\documentclass{aa}

\bibpunct{(}{)}{;}{a}{}{,}
\usepackage{latexsym}
\usepackage{natbib}
\usepackage{graphicx}
\usepackage{subfigure}
\usepackage{txfonts}
\usepackage{color}
\usepackage{booktabs}
\usepackage{multirow}
\usepackage{setspace}
\usepackage[flushleft]{threeparttable}
\usepackage{multicol}
\usepackage{hyperref}
\usepackage{titlesec}
\titlespacing\section{2pt}{12pt plus 4pt minus 2pt}{0pt plus 2pt minus 2pt}
\titlespacing\subsection{2pt}{12pt plus 4pt minus 2pt}{0pt plus 2pt minus 2pt}
\titlespacing\subsubsection{2pt}{12pt plus 4pt minus 2pt}{0pt plus 2pt minus 2pt}

\usepackage{ulem}
\graphicspath{{./}{fig_eps/}{fig_pdf/}}
\begin{document}

%
\title{Case studies of multi-day $^{3}$He-rich solar energetic particle periods} 

\titlerunning{Case studies of multi-day $^{3}$He-rich solar energetic particle events}
\authorrunning{N. H. Chen et al.}

\author{Nai-hwa Chen \inst{1}
\and Radoslav. Bu{\v c}{\'{\i}}k \inst{1}
\and Davina. E. Innes \inst{1} 
\and Glenn. M. Mason \inst{2}}

\institute{Max-Planck-Institut f\"ur Sonnensystemforschung, D-37077 G\"ottingen, Germany 
\and  Applied Physics Laboratory, Johns Hopkins University, Laurel, MD, 27023, USA}

\offprints{N. H. Chen, \email{chenm@mps.mpg.de}}
\date{Received 5 January 2015 ;Accepted }

\abstract%
%
{ Impulsive solar energetic particle events in the inner heliosphere show the long-lasting enrichment of  $^{3}$He. }
{ We study the source regions of long-lasting $^{3}$He-rich solar energetic particle (SEP) events  } 
{We located the responsible open magnetic field regions, we combined potential field source surface extrapolations (PFSS) with the Parker spiral, and compared the magnetic field of the identified source regions with in situ magnetic fields. The candidate open field regions are active region plages. The activity was examined by using extreme ultraviolet (EUV) images from the Solar Dynamics Observatory (SDO) and STEREO together with radio observations from STEREO and WIND.
}
{ Multi-day periods of $^{3}$He-rich SEP events are associated with ion production in single active region. Small flares or coronal jets are their responsible solar sources. We also find that the $^{3}$He enrichment may depend on the occurrence rate of coronal jets.
}
{}
\keywords{%
Sun: corona ---
Sun: UV radiation --- 
Sun: particle emission---
Sun: flares---
} 

\maketitle

\section{Introduction}
{\indent Measurements in the past few decades show that solar energetic particles (SEPs) are accelerated in the shocks of coronal mass ejections (CMEs) and during flares \cite[for a detailed review see in][and the references therein]{reames1999}. The former are generally believed to be responsible for gradual SEP events, and the latter are the likely source of impulsive SEP events \citep{ Cliver1983,Cane1986,Cane1988, kahler1984}. It is known that the impulsive SEP events have a higher $^{3}$He/$^{4}$He isotopic ratio and heavier ion enhancement than in the average corona  \citep{Hsieh1970, Mason1986}. These ions are understood to be accelerated and released above the region where hard X-rays and microwaves are emitted, as inferred from radio observations \citep{ Klein2005}. The high mean ionic charge state of ions found in impulsive SEP events suggests that the temperature of the source region could reach 10$^{7}$ K \citep{klecker1984, luhn1987} in turn suggesting the acceleration site at an altitude less than 0.2 Rs \citep{klecker2007}. Resonant wave-particle interactions in the flare site are thought to be the acceleration mechanisms for $^{3}$He-rich SEP events \citep{Fisk1978, temerin1992}. There have been  several cases reported in which the $^{3}$He flux persisted above the background for a few days, suggesting a steady and continuous acceleration of $^{3}$He ions in their source sites \citep{mason2007book,2008Kocharov}. \\
\indent  In the Solar and Heliospheric Observatory (SOHO) era, it was shown that small brightenings or jets at or near active region (ARs) on the western hemisphere were the source regions of $^{3}$He-rich SEP events as inferred from EUV, white light, and radio observations \citep{ wang2006ApJ639, pick2006,nitta2006}. In their studies, recurrent coronal jets were seen from the connected ARs for a period of 1-2 days around the time of the $^{3}$He-rich SEP event. Coronal jets are often seen when flux emerges into open field regions, which leads to reconnection between the emerging flux and open fields \citep{Yokoyama1996, Moreno-Insertis2008, Archontis2010}. Jets are also commonly observed on the edge of sunspots \citep{1996Canfield,nitta2008a,2011Innes} and in active region plages \citep{wang2006ApJ639, pick2006}. The close temporal correlation between $^{3}$He-rich SEP events, jets, and kilometric type III radio bursts suggests that the energetic particles are accelerated via magnetic reconnection \citep{ Lin1985, Reames1985, reames1986, Kundu1995, Krucker2011, Klassen2011}. \\
  \indent  In this paper, we examine the source regions of SEPs whose $^{3}$He-rich period lasts for 4-5 days. The investigated $^{3}$He-rich SEP events were observed by Advanced Composition Explorer (ACE). In section 2, we investigate the source regions identified with the help of coronal magnetic field extrapolations, and the associated activity on the solar disk seen in solar EUV images of the Atmospheric Imaging Assembly (AIA) on board SDO and the Extreme UltraViolet Imager (EUVI)/STEREO-A. The results are summarized and discussed in section 3.}

 \begin{figure*}[htp] 
 \graphicspath{C:\Users\martinaPC\Desktop\MPS_work\pdfLex\SEP_paper}
 \begin{center}
\subfigure { \includegraphics[width=.45\textwidth]{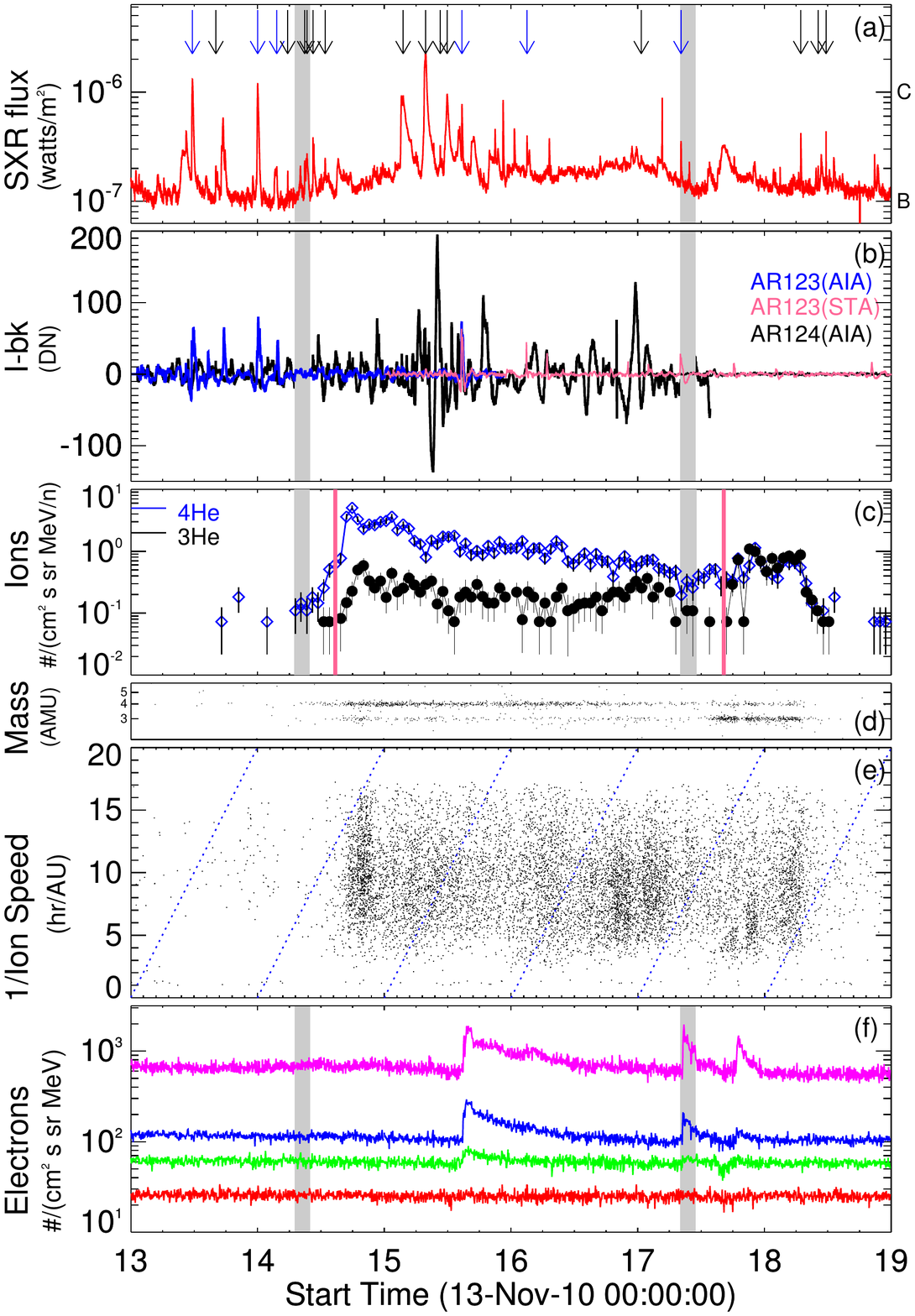}}
 \subfigure{ \includegraphics[width=.465\textwidth]{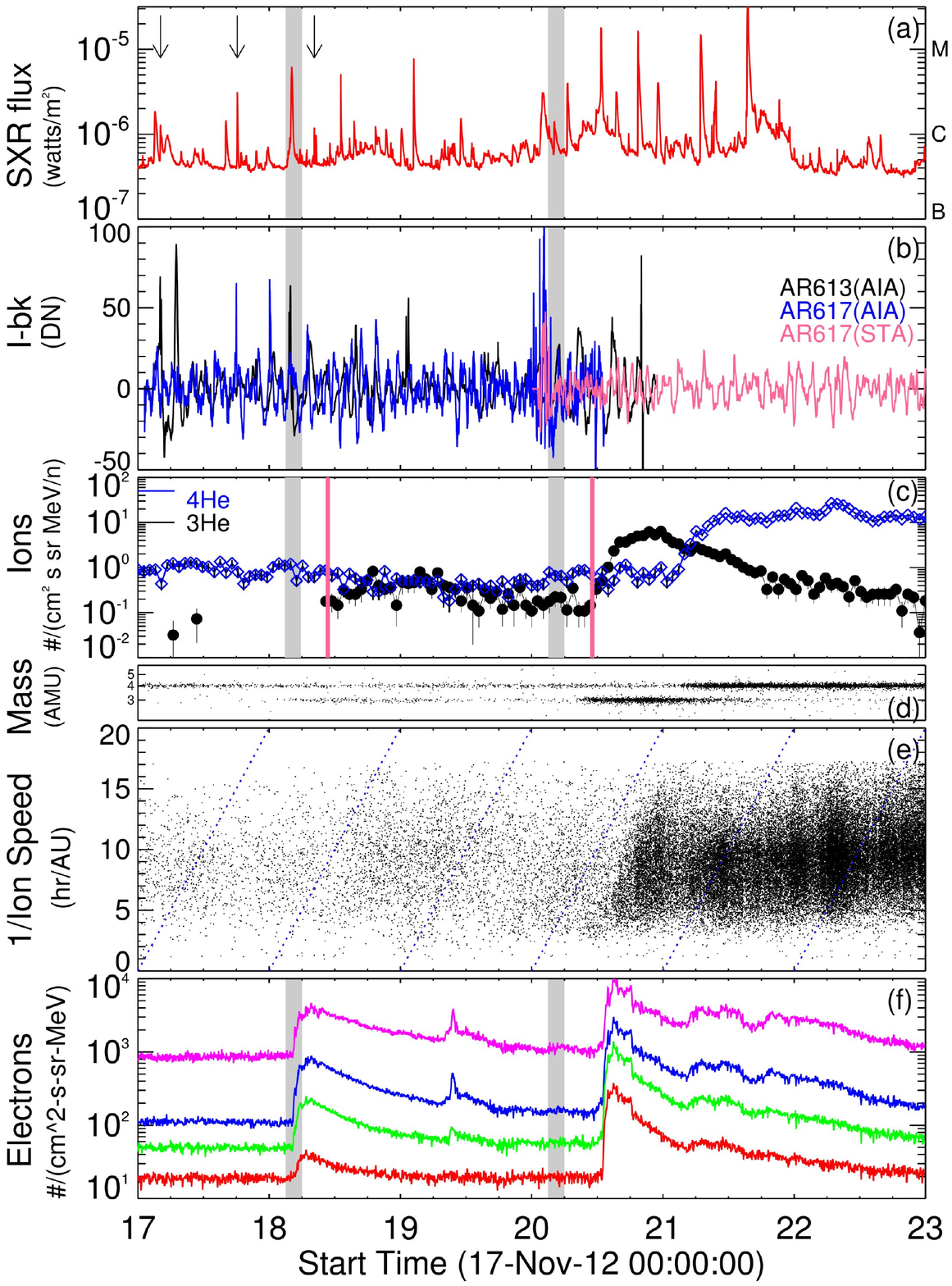}}
  \end{center}
\label {figure 1}
\caption{ Summary of multi-day $^{3}$He periods and related solar activity.  Left: \textbf{Period 1}, Right: \textbf{Period 2}\quad (a) GOES SXR flare flux in the energy band of 1-8 {\AA}. Arrows at the top point to the SXR flares in the candidate ARs and flares in other ARs are not marked. (b) The EUV 193{\AA} excess intensity, the average AR flux (see section 2.1) observed by AIA/SDO and STA in AR11124 (black); 11123 (blue, purple) in period 1 and  AR11613 (black); and 11617 (blue, purple) in period 2. The intensity extension of AR11123 and 11617, which are integrated in EUVI/STA, is shown in purple for two periods. ARs are labeled with the last three digits of their names e.g., AR11124 is labeled as 124. (c) 0.23-0.32 MeV/n 1 hr averaged $^{3}$He (black) and  $^{4}$He (blue) intensity observed by ULEIS/ACE. The purple lines show the approximate start time of each event at 1 AU, shown in Table 1. (d) Helium mass spectrogram in energy range 0.4-10 MeV/n. (e) Inverted ion velocity time spectrogram in mass range 10-70 AMU. (f) The 5 min averaged electron intensity at EPAM/ACE in four energy bands: 0.038- 0.053 MeV( purple), 0.053-0.103 MeV(blue), 0.103-0.175 MeV(green), and 0.175-0.315 MeV(red). The gray regions mark the time window for the $^{3}$He ions release time from the Sun (see section 2.1). }
\end{figure*}

\begin{table*}[tbp]
 \centering
\resizebox{0.5\textwidth}{!}{
\begin{threeparttable}
\caption{$^{3} $He-rich SEP events in this study}
\begin{tabular}{lllc} 
\toprule[1.2pt] 
 event & start time (UT)& end time (UT) &  $^{3}$He/$^{4}$He (ACE) \\ \midrule
1 &2010-Nov-14  15:00 & 2010-Nov-17 09:00& 0.2  \\
    &2010-Nov-17  16:00& 2010-Nov-18 10:00 & 2.8  \\ \midrule
2 &2012-Nov-18  11:00& 2012-Nov-20 06:00 & 1.0 \\
   &2012-Nov-20  11:00& 2012-Nov-21 04:00 & 6.7  \\  \bottomrule[1.2pt]
\end{tabular}
 \tablefoot{The ion event starting time is inferred from $^{3}$He intensity time profiles in energy of 0.23-0.32 MeV/n and the ratios are from energy of 0.32 -0.45 MeV/n.}
 \end{threeparttable}}
\end{table*}

\begin{table*}[tbp]
\centering
\caption{GOES SXR Flares and Type III/VI bursts around the estimated time windows.}
\resizebox{0.9\textwidth}{!}{
\begin{tabular}{lccccl} 
\toprule[1.1pt] 
 period & date & flare start /peak time (UT) & GOES class & Active Region & Type III/VI bursts Time \\   \midrule
1  &2010-Nov-13 & 23:50/00:01&C1.0&11123& 23:40(-00:02) (type VI)  \\
  &2010-Nov-14 &03:08/03:38&B2.0&11123&\\  
  & &08:52/08:55&B2.2&11124&\\  
    & &09:08/09:22&B2.6&11124&\\  
    &2010-Nov-17&  08:07/08:12&B3.4&11123& 08:08(-08:10) \\     \midrule
 2  &2012-Nov-18 &  &  & & 00:06-00:09 \\  
   & &  &  & & 03:27-03:30\\
     & &  &  &11615 & 03:46(-04:13) (type VI) \\ 
    & &  03:55/ 04:07&C5.7&11615& 04:00(-04:07) \\
       & &  &  & & 07:28(-07:29) \\
   & 2012-Nov-19 &  &  & & 23:40 \\
      & 2012-Nov-20 &  &  & & 00:18 \\
      & &  &  & & 01:29 \\
    & & 01:37/ 02:07 & C3.0 & 11618& 02:10  \\
      & &  &  & \textbf{11617}& 02:32(-02:39) (type VI) \\
    & &  &  & & 03:09 \\
      & &  &  & & 03:29 \\
  &  & 04:04/ 04:10&C1.3&11618&\\   \bottomrule[1.1pt]
\end{tabular}}
 \tablefoot {The bold AR11617 in 2012-Nov-20 shows the corrected source of the type III burst that we identified (see section 2.1.2). 
 }
\end{table*}

\section{Observation and data analysis} 
\subsection{$^{3}$He-rich SEP events}{\indent  We investigate the solar sources of two long-lasting $^{3}$He-rich periods on 2010 November 14-18 and 2012 November 18-21 (Figure 1(c)). Each consists of two individual events. The helium ions were detected by the Ultra Low Energy Isotope Spectrometer \citep[ULEIS]{Mason1998} on ACE at the L1 Lagrangian point. We first estimate the release time window of the $^{3}$He-rich SEP events based on the ion energy of 0.23-0.32 MeV nucleon$^{-1}$. Under the assumption that these particles travel scatter-free along the spiral path length of 1.2 AU, it is about 5-8 hours prior to event start time for events showing energy dispersion in the inverse velocity-time spectrograms. The release time for dispersionless events is difficult to estimate because the spacecraft connects to field lines previously populated with SEPs. Table 1 contains the approximate start/end times of the ion events at 1 AU and events summed the $^{3}$He/$^{4}$He ratios \citep{2013Bucik}. We list the flares that occurred close to the release time windows and the related type III and type VI bursts in Table 2. These are based on the Space Weather Prediction Center (SWPC) daily solar event report. Type VI are a series of type III bursts over a period of 10 minutes or more. The activity at the Sun during the $^{3}$He-rich SEP periods is summarized as soft X-ray and EUV light curves in Figure 1. Here we show in panel (a) GOES soft X-ray (SXR) flux (1-8 {\AA}), and in (b) the spatially-averaged EUV 193{\AA} flux of associated source regions, discussed further in section 2.3. The EUV flux is integrated over the dashed boxes (see fig. 4) in the AIA observations with the 30-minute average background subtracted. The STA observations are also added when the ARs cross the limb. The background is possibly influenced by the duration and the occurrence rate of flares. Negative intensity is sometimes caused by a large background, which is due to the overlap of two close flares. Also shown are (c) $^{3}$He and $^{4}$He ions flux observed by ULEIS/ACE, (d) helium mass spectrogram, (e) inverted ion velocity time spectrogram, and (f) 5-minute averaged electron intensity in EPAM/ACE in the energy channels between 38-315 keV.  
\subsubsection{ 2010 November 14-18}{\indent Period 1, shown in the left panels of Figure 1, consists of two individual $^{3}$He flux increases, one starting on Nov. 14 and the other on Nov. 17. The first increase on Nov. 14 showed an $^{3}$He/$^{4}$He ratio of $\sim$0.20 (0.32-0.45 MeV nucleon$^{-1}$) but no accompanying electron event was observed. The possible associated SXR flares in the estimated time window were two B-class flares from AR11124 with start time at 08:52 and 09:08 UT. We note that these were not associated with type III emission.  The November 14 event shows a dispersionless onset, where ions of different speeds enhance simultaneously. It occurs when the spacecraft encounters a magnetic flux tube already filled with energetic ions. The ion release time most likely occurred before the indicated time window. The second increase on November 17 had a higher $^{3}$He/$^{4}$He ratio of $\sim$2.75. It was associated with a B3.4 SXR flare in AR11123, which peaked at 08:12 UT. It was accompanied by an impulsive electron event and the ion velocity dispersion was clearly evident (indicated by the typical triangular pattern). We also searched for type III radio bursts in WIND/WAVES data that occur close to the release time window, shown in Figure 2. The red bars mark the estimated release time windows. An expanded view of panel (a) around the time of the red bars is given in panels (b) and (c) below. Since there was no electron event associated with the first $^{3}$He increase, it is not surprising that there was no type III burst found in our time window. Earlier VI/III bursts were observed at 23:40 UT, Nov. 13, and 03:08 UT, Nov. 14. These two bursts were accompanied by C1.0 and B2.0 flares, respectively, in AR11123 (see Table 2). In the second $^{3}$He increase, a type III burst with an onset at 08:10 UT was clearly observed and was accompanied by a B3.4 flare in AR11123.}
\subsubsection{2012 November 18-21}  {\indent The corresponding plots for period 2 are shown in the right panels of Figure 1. The energetic $^{3}$He flux increase began on Nov. 18 with another enhancement observed on Nov. 20 (panel (c)). The first increase, with $^{3}$He/$^{4}$He ratio of $\sim$1.0, was accompanied by a C5.7 SXR flare from AR11615 and an electron event with a longer duration.  The $^{3}$He-rich event has no clear velocity dispersion. The radio observations in the right panel (b) of Figure 2 showed a type III/VI burst associated with the C5.7 flare, starting at 03:46 UT in the release time window. In addition, the  0.4-10 MeV/n $^{3}$He was already present at the flare start time (see the mass spectrogram panel (d)). Therefore this event probably originated from an earlier release, but the contribution of $^{3}$He later during the event cannot be ruled out. There was a type III burst with onset at 00:06 UT which is under our consideration for the possible earlier release. This has no associated source region in the SWPC report. The second increase showed a $^{3}$He/$^{4}$He ratio of 6.7, but no corresponding electron event was observed. The prominent electron event later was related to the M1.7 SXR flare with the onset time at 12:36 UT. The $^{3}$He-rich event velocity dispersion was clearly evident. The two C-class SXR flares that occurred close to the time window (Table 2) were from AR11618, located in the eastern hemisphere. Five type III/VI bursts were observed between 01:00 and 04:00 UT, Nov. 20. Among them, two bursts (marked by arrows in the right panel (c) in Figure 2) with onset at 01:29 and 02:32 were quiet strong, showing a deep drift below 0.1 MHz. The source AR for the 01:29 UT burst was not reported and for the 02:32 UT the source in the SWPC list, AR11618, was likely erroneous. AR11618 (E27) was located near the west limb (W98) in the STEREO-B (STB) view and thus not far from the spacecraft magnetic connection point, but no type III burst was seen by the WAVES/STB radio instrument. However, STA observed the same 01:29, 02:32 type III bursts even though the purported source AR11618 was at the back side of the Sun (E155) from the STA view. This makes it very unlikely that AR11618 was the source of these two type III bursts. On the other hand, AR11617 shows a clear EUV brightening at the same time as type III burst at 02:32 UT (Sec. 2.3). Additionally, a weak narrow coronal mass ejections (CME), observed by LASCO (Large Angle and Spectrometric Coronagraph)/SOHO, lifted off from the west limb at the time between 2:30 and 4:00 UT after the brightening seen on AR11617. The lack of the associated electron event might be due to the observational effect. The electron intensity level is slightly higher due to the previous two electron events which might mask the related small event. We note that the previous studies reported only $\sim$ 60{\%} association with the electron events (e.g., \citep{nitta2006}) } \\[2 pt]
\indent For the two periods, not all the investigated $^{3}$He increases are related to electron events or show velocity dispersion. Thus their origins are not clearly evident in the energetic ion temporal profiles, SXR flare, and radio observations. To further narrow down the possible source regions, we examine the magnetic field connection and EUV images in the next sections.}
\begin{figure*}[htp] 
\graphicspath{C:\Users\martinaPC\Desktop\MPS_work\pdfLex\SEP_paper}
\begin{center}
  \subfigure { \includegraphics[width=.45\textwidth]{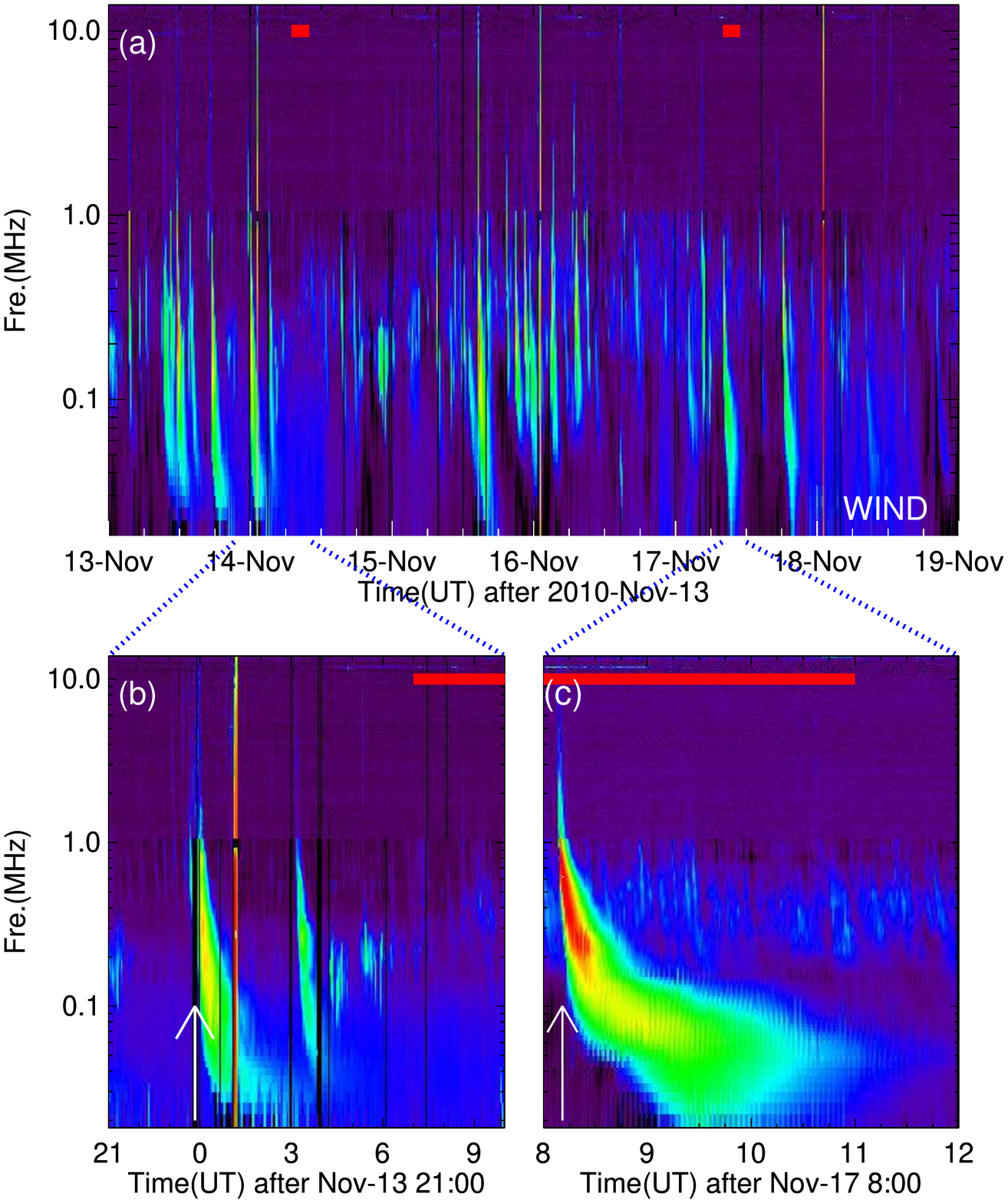}}
 \subfigure{ \includegraphics[width=.45\textwidth]{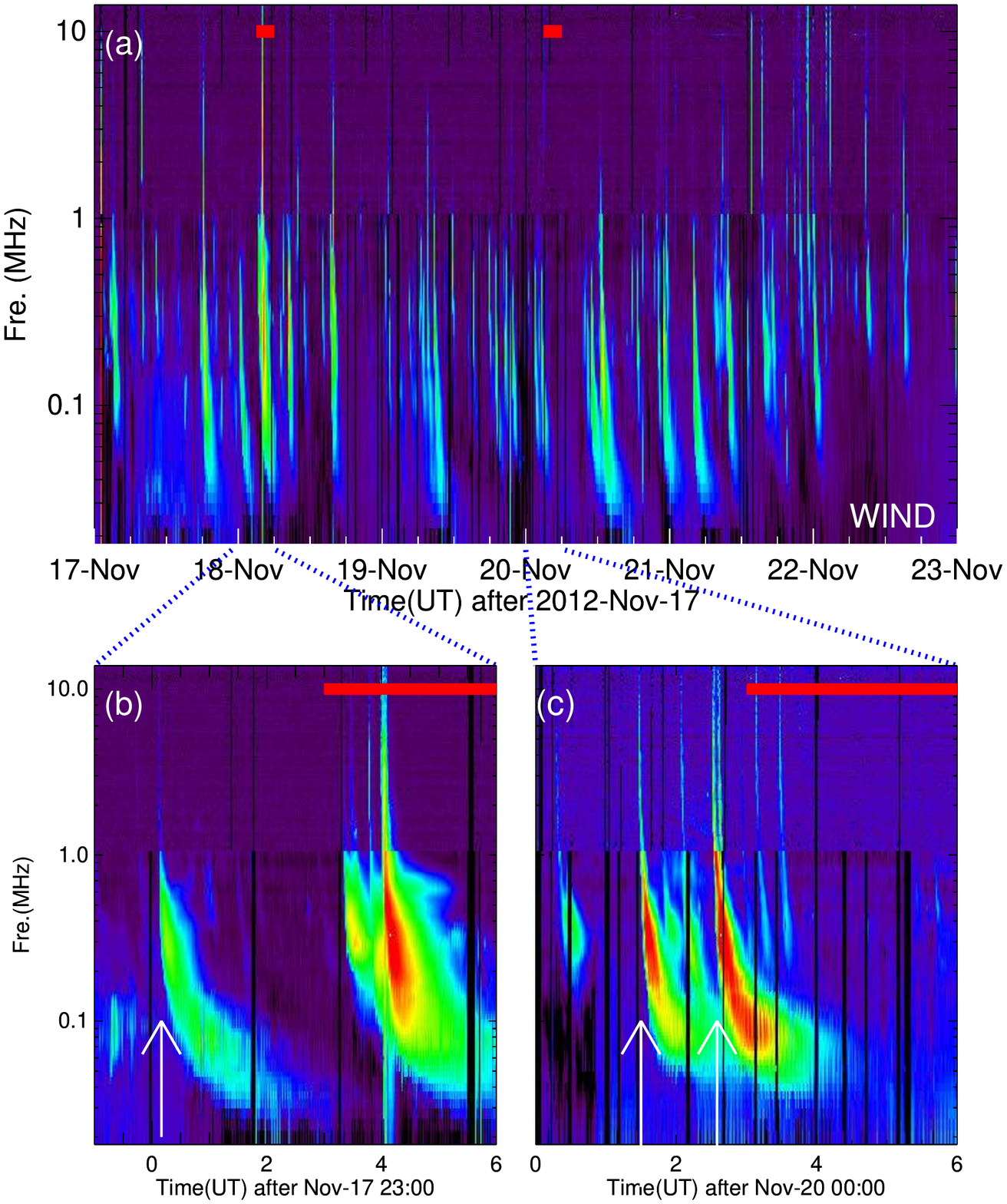}}
 \end{center}
 \label {figure 2}
 \caption{Kilometric radio observations. (left): \textbf{period 1};(right): \textbf{period 2}\quad  Panel (a): Kilometric radio observations on WIND/WAVE in two events. The red bars mark the ion release time windows shown as the dashed regions in Figure 1.  Panels (b) and (c) are enlarged views of the time ranges around each event associated type III radio burst, indicated by white arrows.}
\end{figure*}

\begin{figure*}[htp] 
\begin{center}
 \subfigure{  \includegraphics[width=.45\textwidth]{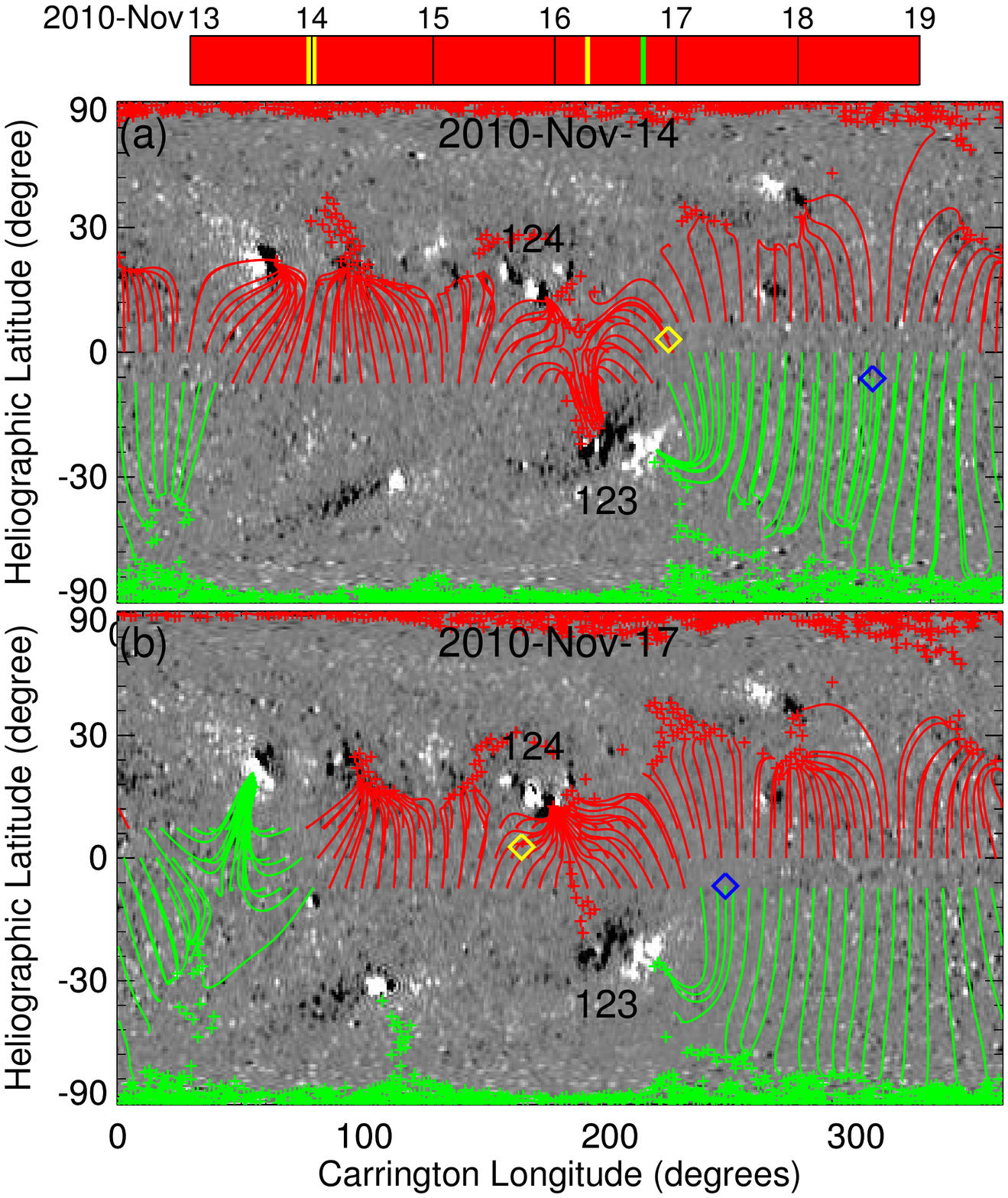}}
\subfigure{ \includegraphics[width=.45\textwidth]{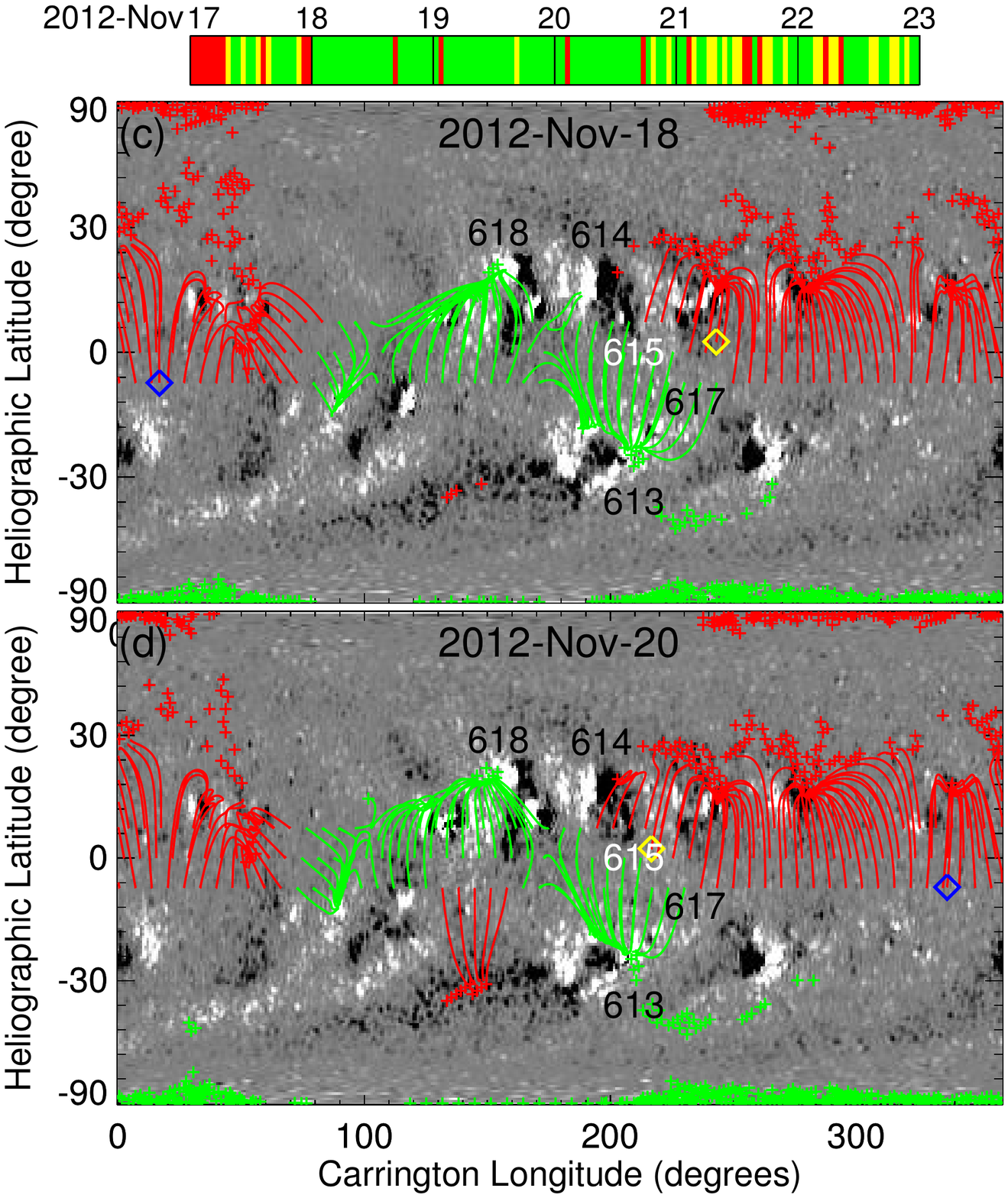}}
 
 \end{center}
 \label {Figure 3}
 \caption{ The photospheric magnetic field synoptic map with coronal magnetic field lines derived from the PFSS model at the times of the $^{3}$He ions flux increases. The green and red lines represent the positive and negative polarity open field lines that intersect the source surface at latitudes $ 0^{\circ}$ and $\pm7^{\circ}$. Field lines that are open at the ecliptic are shown by lines and other regions of open field are denoted as crosses. The color bars above indicate the interplanetary magnetic field polarity. The yellow and blue diamonds show the footpoint of Earth and STA on the source surface, respectively. We note that the data source of the map is based on HMI magnetograms (grey scale) and the time runs from right to left in the map. ARs are labeled with the last three digits of their names e.g., AR11124 is marked at 124. }
\end{figure*}

 \begin{figure*}[htp] 
\begin{center}
  \subfigure { \includegraphics[width=.45\textwidth]{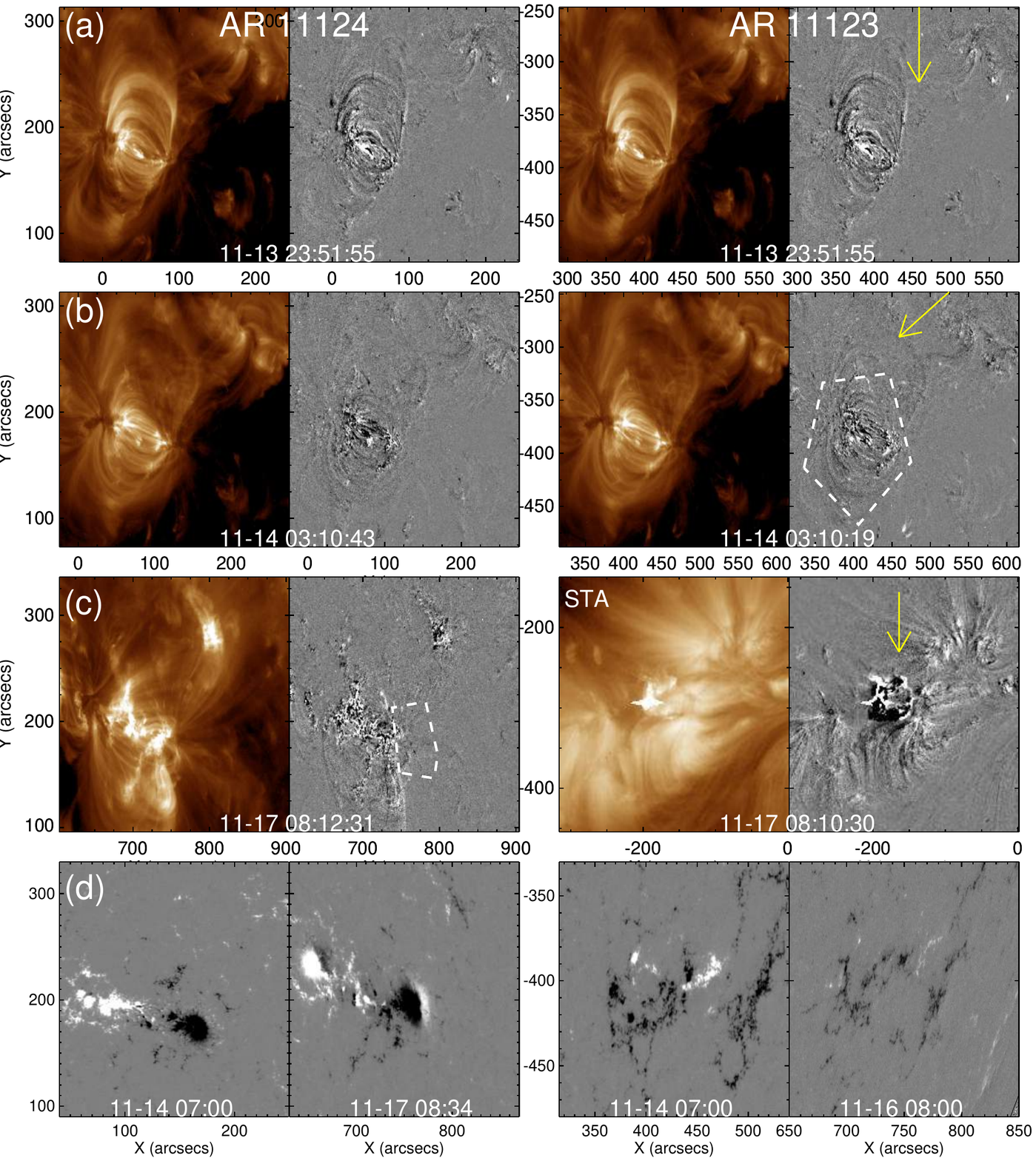}}
 \subfigure { \includegraphics[width=.45\textwidth]{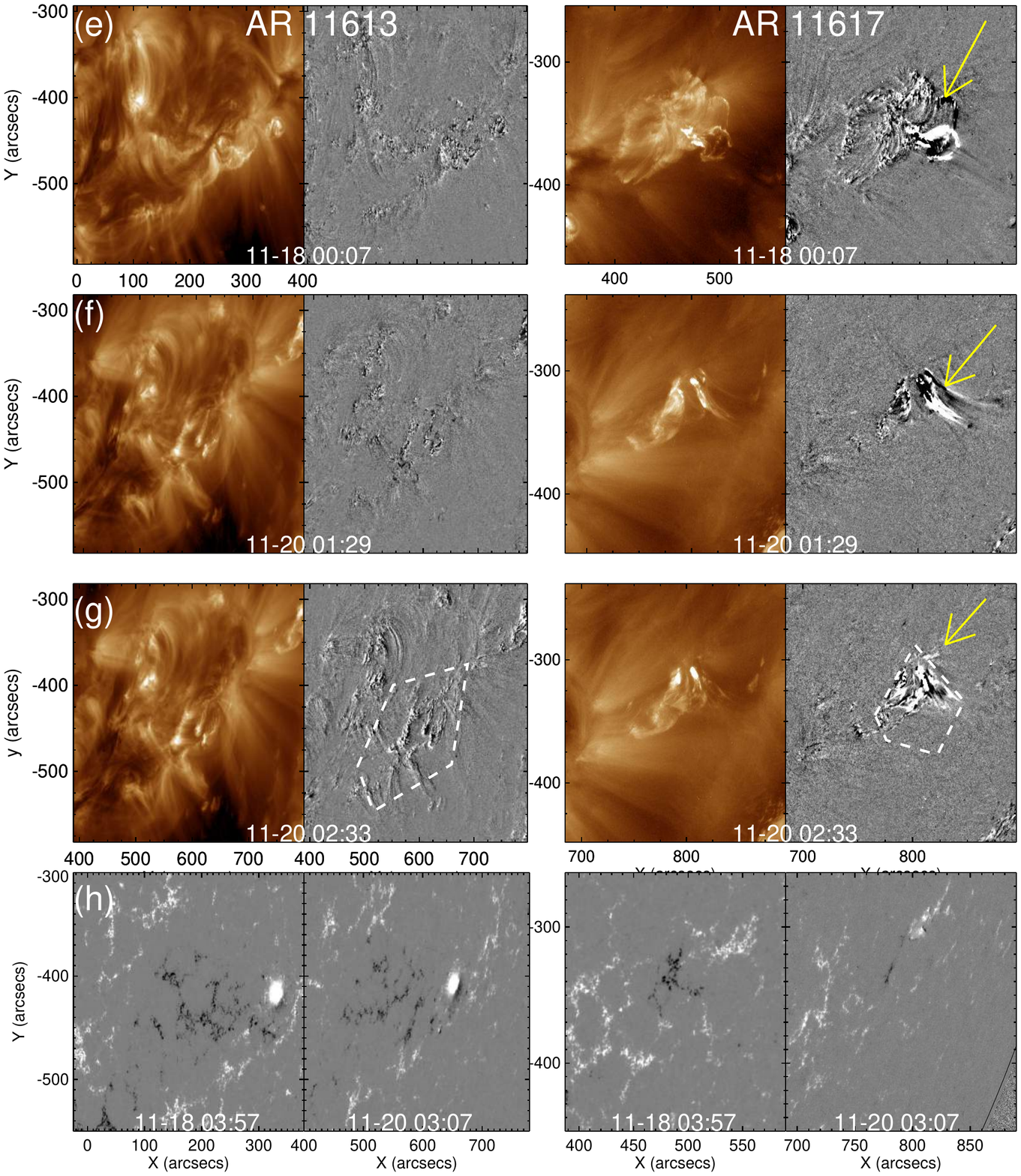}}
 \caption{AIA observations on the related source regions. a-d: \textbf{period 1-AR11123, 11124}, e-h: \textbf{period 2-AR11613, 11617}\quad The AIA 193{\AA}intensity/running difference images in the candidate ARs at times of type III radio bursts in two periods. The flaring/jets are indicated by the yellow arrows. Panels d and h are the HMI magnetograms of each candidate at the beginning time of each release window. For AR11123, nearly invisible at Nov. 17, we choose Nov. 16 instead. The dashed boxes are the integrated regions for the EUV light curves in Figure 1b.}
\end{center}
\end{figure*}
 \subsection{Magnetically connected regions}
 {\indent To find the source regions of the investigated $^{3}$He-rich SEP events, we determine the connectivity from the Sun through the corona and interplanetary space to Earth by the method combining the potential field source surface (PFSS) model and Parker spiral. It has been used to identify the sources of $^{3}$He-rich SEP events in previous investigations \citep{nitta2006, wang2006ApJ639}. The data source used for the PFSS model here, which is available via Solar Soft PFSS package (\url{http://www.lmsal.com/_derosa/pfsspack}), is based on the SDO Helioseimic and Magnetic Imager (HMI) magnetograms.  In addition, the in situ magnetic polarity is compared with the results of PFSS model in this study. We trace the field lines back from the source surface at 2.5 $R_{sun}$ to the photosphere using the PFSS model \citep{schrijver2003}. The interplanetary magnetic field between the source surface and the spacecraft, assumed as a Parker spiral is a function of the measured solar wind speed. In Figure 3, the open field lines are overplotted on the photospheric magnetic maps that were reconstructed at the start times (see Table 1) of the $^{3}$He enrichments. The green and red lines represent the positive and negative polarities, respectively. The color-bar above the map indicates the in situ magnetic polarity during the time of $^{3}$He periods with the same color index as the field line tracing (e.g. green indicates the magnetic field pointing away from the Sun). The in situ polarity is determined by the angle between the observed magnetic field and the Parker spiral model magnetic field using the measured solar wind speed on ACE. When the angle is within $\pm20^{\circ}$, of the normal Parker field direction, yellow is used.  \\     
 \indent  Figures 3a and 3b show magnetic connections for the 2010 November 14 and 17 events, respectively. The in situ magnetic polarity suggests that on November 14, ACE was connected via negative (red) field to AR11123 or to the coronal hole located below AR11124.  The PFSS extrapolations for November 17 suggests the connection to AR11124. However, at this time, AR11123 was very close to the west limb so the magnetic field was poorly measured and the PFSS extrapolation may not reflect the field as it was. We therefore require EUV data to identify the source region.\\
 \indent The corresponding maps for the 2012 November 18 and 20 events are shown in Figures 3c and 3d. Multiple regions on the solar surface revealed the complexity during this $^{3}$He-rich period. It was also seen that the in situ magnetic polarity changed several times to the negative or uncertain (yellow) polarity, but the main connection was via the positive field (green). The in situ polarity suggests its connection through the positive field to the area adjacent to AR11617 and AR11613 for both November 18 and 20. }
 \subsection{EUV observations in the associated  active regions}{\indent We use the AIA 193{\AA} 1 min and EUVI/STA 5 min 195{\AA} image observations to study the EUV variation in the candidate ARs during the investigated periods. The 193{\AA} channel primarily centered on Fe XII line is sensitive to the warm plasma in the vicinity of ARs at 1.6 MK and secondarily (Fe XXIV) to 18 MK plasma. Figure 4 shows images of the candidate ARs for the two periods. The dashed boxes mark the regions of interest over which the ARs light curves are integrated. The selected boxes are centered on the regions with the conspicuous appearance of jet-like emissions or flaring in the main loops. We also show an STA observation of AR11123 when it was at the west limb in the AIA images. The bottom panels are the HMI magnetograms of each candidate.  \\[7pt] 
 \textsl{2.3.1.} \textbf{2010 November 14-18}\\[3pt]
 \indent The EUV intensity and running-difference images of the candidates AR11123 and 11124 are shown in Figure 4 at the onset of associated type III radio bursts. In the background-subtracted EUV light curve of two candidate ARs (Figure 1b (left)), the EUV spikes in AR11123 were stronger than those generated by AR11124 in the early two days and then activity on AR11124 increased at the rest of time. Those EUV spikes in AR11124 were attributed to the jet-like emissions and AR11123 was in favor of flaring. For the two ion release time windows, clear EUV spikes, peaked at Nov. 14 00:01 03:38 UT and Nov. 17 08:12 UT (C1.0, B2.0, and B3.4 flares), coincided with two type III/VI bursts on November 14 (Figure 4a, 4b) and the type III burst on November 17 (Figure 4c) were observed in AR11123, but no simultaneous brightening appeared in AR11124.  During the multiday event starting on Nov. 14, an EUV spike (B7.6 flare) at Nov. 15 14:38 UT with an associated electron event and a simultaneous type III burst also originated in AR11123. Throughout the whole period, small flares emanated from AR11123 and could contribute to the observed $^{3}$He enrichment. \\[7pt]
\textsl{ 2.3.2} \textbf{2012 November 18-21}\\[3pt]
\indent The corresponding EUV images of the two nearby candidates AR11613 and 11617, are shown in Figure 4e-g at the onset of the associated type III radio bursts. Their intensity integrations (Figure 1b (right)) showed numerous spikes produced by jets during this period. In the first enrichment, an intense EUV spike of AR11617 in EUV light curve was observed contemporarily with type III burst on November 18 00:06 UT (also seen in Figure 4e). The second enrichment, seen on November 20, was preceded by an increase in jet activity from AR11617 (see Figure 1b (right)). Among them, two distinct jets (01:30, 02:33 UT) from AR11617 were clearly evident in the AIA images coinciding with two type III/VI bursts on November 20 (see Figures 4f and 4g). These jets were clearly observed shortly after the type III burst and before the $^{3}$He abundance increase. Thus, the frequent jets from AR11617 are associated with the energetic $^{3}$He production in this period.\\
\indent The magnetograms, at the beginning time of each release window in two periods, show where jets or flares in our candidates occurred. The jet-produced area of AR11124 originated from the west edge of a sunspot, shown in the panel (d), while AR11123 was an AR plage. AR11617, where numerous jets were produced, was also an AR plage (see Figure 4h).\\
\indent The AIA/SDO observations have much higher cadence and spatial resolution than the EUV Imaging Telescope in the previous works \citep{wang2006ApJ639,nitta2006}, which suggests the reasonable sources in the temporal investigation. The HMI magnetograms show that the activity of our candidate sources emanate from a plage region. Our observations provide evidence that the flaring/jet on a single AR is responsible for the $^{3}$He abundance increase during the whole period.}
\section{Discussion and summary}
{\indent We have studied the solar origins of two long-lasting $^{3}$He-rich SEP periods by analyzing high cadence coronal images. The source regions during the two periods, determined by the PFSS extrapolations of photospheric magnetic field together with EUV and radio observations, appear to be single ARs with a long steady connection to the observer. The source regions of two huge $^{3}$He abundance events ($^{3}$He/$^{4}$He >1) produce flares/jets with correlated type III/VI bursts. These impulsive SEP events show clear velocity dispersion onsets. For the two dispersionless events, the timing of type III bursts are used to determine the possible sources due to their high association with 3He-rich SEP event \citep{Reames1985, nitta2006}. \\ 
\indent The 2010 November 14 and 17 events were associated with AR11123 when it was at W26-W28 and near the west limb (W91), respectively. The 2012 November 18 and 20 events were associated with AR11617 when it was at W33 and W70, respectively. Previous single spacecraft observations have reported multiple events over a two-day period \citep{reames1986, wang2006ApJ639,2000Mason}. The coronal field extrapolations in our candidate regions revealed substantial longitudinal spread of the ecliptic field lines allowing connection for a longer period. Multiple observations with widely separated spacecraft have suggested that ARs may produce 3He-rich SEP events even for a quarter of a solar rotation \citep{buick2014}. It has also been suggested that multiday periods of 3He-rich SEPs may be due to confinement of the ions in the co-rotating interaction regions in the solar wind \citep{2008Kocharov}. \citet{mason2007book} has suggested a continuous injection from the solar sources when onset was not obvious. \\
\indent We have shown two multi-day periods associated with the same source where the first events show lower $^{3}$He-enrichment than the second ones. In the 2012 November 18-21 period, numerous recurrent EUV jets occur in AR11617, an AR plage, for six hours (from the November 19 22:00 UT to November 20 4:00 UT) preceding the second event. Their magnitude and especially the frequency of occurrence largely increases. A type VI burst accompanied by these jets suggests the continuous accelerations and ejections of low-energy electrons from the source site for a longer time than single normal type III burst does. It might suggest that huge $^{3}$He abundance in the second event is due to the continuous processes in its source. In the 2010 November 14-18 period, the responsible activity is flaring instead of the jetting. Source AR11123 has generated many small flares since its maximum phase (on November 11, see \citet{2014Mandrini}) and decay during our investigated period. The sizes of two responsible flares are different by one order of magnitude. Stronger $^{3}$He enrichment in later time is associated with the smaller flares (B3.4) which is consistent with previous works \citep{reames1988b,reames1999}. \\
 \indent Our observations provide evidence that the production and release of $^{3}$He-rich SEPs from a single AR may be a common process. Small flares or coronal jets, noted in previous works, appear to be the solar source resposible for the $^{3}$He-rich SEP in this study. The $^{3}$He enrichment might also depend on the occurrence rate of coronal jets and the duration of ejection time seen in our observations.
}

\begin{acknowledgement}
\small
{We thank the SDO, WIND/WAVES, ACE and STEREO team for the data. This work was supported by National Science Council of Taiwan under grant NSC 103-2917-I-564-007 and the Postdoctoral Research Abroad Program, and by NASA under grant NNX13AR20G/115828. }
\end{acknowledgement}

\bibliographystyle{aa} 
 \bibliography{main_draft} 

\begin{thebibliography}{37}
\expandafter\ifx\csname natexlab\endcsname\relax\def\natexlab#1{#1}\fi

\bibitem[{{Archontis} {et~al.}(2010){Archontis}, {Tsinganos}, \&
  {Gontikakis}}]{Archontis2010}
{Archontis}, V., {Tsinganos}, K., \& {Gontikakis}, C. 2010, \aap, 512, L2

\bibitem[{{Bucik} {et~al.}(2013){Bucik}, {Innes}, {Mall}, {Korth}, \&
  {Mason}}]{2013Bucik}
{Bucik}, R., {Innes}, D.~E., {Mall}, U., {Korth}, A., \& {Mason}, G.~M. 2013,
  Proc. 33rd ICRC (Rio de Janeiro), paper SH-EX 0552 (arXiv:1307.6342)

\bibitem[{{Bu{\v c}{\'{\i}}k} {et~al.}(2014){Bu{\v c}{\'{\i}}k}, {Innes},
  {Mall}, {Korth}, {Mason}, \& {G{\'o}mez-Herrero}}]{buick2014}
{Bu{\v c}{\'{\i}}k}, R., {Innes}, D.~E., {Mall}, U., {et~al.} 2014, \apj, 786,
  71

\bibitem[{{Cane} {et~al.}(1986){Cane}, {McGuire}, \& {von
  Rosenvinge}}]{Cane1986}
{Cane}, H.~V., {McGuire}, R.~E., \& {von Rosenvinge}, T.~T. 1986, \apj, 301,
  448

\bibitem[{{Cane} {et~al.}(1988){Cane}, {Reames}, \& {von
  Rosenvinge}}]{Cane1988}
{Cane}, H.~V., {Reames}, D.~V., \& {von Rosenvinge}, T.~T. 1988, \jgr, 93, 9555

\bibitem[{{Canfield} {et~al.}(1996){Canfield}, {Reardon}, {Leka}, {Shibata},
  {Yokoyama}, \& {Shimojo}}]{1996Canfield}
{Canfield}, R.~C., {Reardon}, K.~P., {Leka}, K.~D., {et~al.} 1996, \apj, 464,
  1016

\bibitem[{{Cliver} {et~al.}(1983){Cliver}, {Kahler}, \&
  {McIntosh}}]{Cliver1983}
{Cliver}, E.~W., {Kahler}, S.~W., \& {McIntosh}, P.~S. 1983, \apj, 264, 699

\bibitem[{{Fisk}(1978)}]{Fisk1978}
{Fisk}, L.~A. 1978, \apj, 224, 1048

\bibitem[{{Hsieh} \& {Simpson}(1970)}]{Hsieh1970}
{Hsieh}, K.~C. \& {Simpson}, J.~A. 1970, \apjl, 162, L191

\bibitem[{{Innes} {et~al.}(2011){Innes}, {Cameron}, \& {Solanki}}]{2011Innes}
{Innes}, D.~E., {Cameron}, R.~H., \& {Solanki}, S.~K. 2011, \aap, 531, L13

\bibitem[{{Kahler} {et~al.}(1984){Kahler}, {Sheeley}, {Howard}, {Michels},
  {Koomen}, {McGuire}, {von Rosenvinge}, \& {Reames}}]{kahler1984}
{Kahler}, S.~W., {Sheeley}, Jr., N.~R., {Howard}, R.~A., {et~al.} 1984, \jgr,
  89, 9683

\bibitem[{{Klassen} {et~al.}(2011){Klassen}, {G{\'o}mez-Herrero}, \&
  {Heber}}]{Klassen2011}
{Klassen}, A., {G{\'o}mez-Herrero}, R., \& {Heber}, B. 2011, \solphys, 273, 413

\bibitem[{{Klecker} {et~al.}(1984){Klecker}, {Hovestadt}, {Scholer},
  {Gloeckler}, {Ipavich}, {Fan}, \& {Fisk}}]{klecker1984}
{Klecker}, B., {Hovestadt}, D., {Scholer}, M., {et~al.} 1984, \apj, 281, 458

\bibitem[{{Klecker} {et~al.}(2007){Klecker}, {M{\"o}bius}, \&
  {Popecki}}]{klecker2007}
{Klecker}, B., {M{\"o}bius}, E., \& {Popecki}, M.~A. 2007, \ssr, 130, 273

\bibitem[{{Klein} \& {Posner}(2005)}]{Klein2005}
{Klein}, K.-L. \& {Posner}, A. 2005, \aap, 438, 1029

\bibitem[{{Kocharov} {et~al.}(2008){Kocharov}, {Laivola}, {Mason}, {Didkovsky},
  \& {Judge}}]{2008Kocharov}
{Kocharov}, L., {Laivola}, J., {Mason}, G.~M., {Didkovsky}, L., \& {Judge},
  D.~L. 2008, \apjs, 176, 497

\bibitem[{{Krucker} {et~al.}(2011){Krucker}, {Kontar}, {Christe}, {Glesener},
  \& {Lin}}]{Krucker2011}
{Krucker}, S., {Kontar}, E.~P., {Christe}, S., {Glesener}, L., \& {Lin}, R.~P.
  2011, \apj, 742, 82

\bibitem[{{Kundu} {et~al.}(1995){Kundu}, {Raulin}, {Nitta}, {Hudson},
  {Shimojo}, {Shibata}, \& {Raoult}}]{Kundu1995}
{Kundu}, M.~R., {Raulin}, J.~P., {Nitta}, N., {et~al.} 1995, \apjl, 447, L135

\bibitem[{{Lin}(1985)}]{Lin1985}
{Lin}, R.~P. 1985, \solphys, 100, 537

\bibitem[{{Luhn} {et~al.}(1987){Luhn}, {Klecker}, {Hovestadt}, \&
  {Moebius}}]{luhn1987}
{Luhn}, A., {Klecker}, B., {Hovestadt}, D., \& {Moebius}, E. 1987, \apj, 317,
  951

\bibitem[{{Mandrini} {et~al.}(2014){Mandrini}, {Schmieder}, {D{\'e}moulin},
  {Guo}, \& {Cristiani}}]{2014Mandrini}
{Mandrini}, C.~H., {Schmieder}, B., {D{\'e}moulin}, P., {Guo}, Y., \&
  {Cristiani}, G.~D. 2014, \solphys, 289, 2041

\bibitem[{{Mason}(2007)}]{mason2007book}
{Mason}, G.~M. 2007, {$^{3}$He-Rich Solar Energetic Particle Events}, ed.
  R.~{von Steiger}, G.~{Gloeckler}, \& G.~M. {Mason} (Springer Science+Business
  Media), 231

\bibitem[{{Mason} {et~al.}(2000){Mason}, {Dwyer}, \& {Mazur}}]{2000Mason}
{Mason}, G.~M., {Dwyer}, J.~R., \& {Mazur}, J.~E. 2000, \apjl, 545, L157

\bibitem[{{Mason} {et~al.}(1998){Mason}, {Gold}, {Krimigis}, {Mazur},
  {Andrews}, {Daley}, {Dwyer}, {Heuerman}, {James}, {Kennedy}, {Lefevere},
  {Malcolm}, {Tossman}, \& {Walpole}}]{Mason1998}
{Mason}, G.~M., {Gold}, R.~E., {Krimigis}, S.~M., {et~al.} 1998, \ssr, 86, 409

\bibitem[{{Mason} {et~al.}(1986){Mason}, {Reames}, {von Rosenvinge}, {Klecker},
  \& {Hovestadt}}]{Mason1986}
{Mason}, G.~M., {Reames}, D.~V., {von Rosenvinge}, T.~T., {Klecker}, B., \&
  {Hovestadt}, D. 1986, \apj, 303, 849

\bibitem[{{Moreno-Insertis} {et~al.}(2008){Moreno-Insertis}, {Galsgaard}, \&
  {Ugarte-Urra}}]{Moreno-Insertis2008}
{Moreno-Insertis}, F., {Galsgaard}, K., \& {Ugarte-Urra}, I. 2008, \apjl, 673,
  L211

\bibitem[{{Nitta} \& {De Rosa}(2008)}]{nitta2008a}
{Nitta}, N.~V. \& {De Rosa}, M.~L. 2008, \apjl, 673, L207

\bibitem[{{Nitta} {et~al.}(2006){Nitta}, {Reames}, {De Rosa}, {Liu}, {Yashiro},
  \& {Gopalswamy}}]{nitta2006}
{Nitta}, N.~V., {Reames}, D.~V., {De Rosa}, M.~L., {et~al.} 2006, \apj, 650,
  438

\bibitem[{{Pick} {et~al.}(2006){Pick}, {Mason}, {Wang}, {Tan}, \&
  {Wang}}]{pick2006}
{Pick}, M., {Mason}, G.~M., {Wang}, Y.-M., {Tan}, C., \& {Wang}, L. 2006, \apj,
  648, 1247

\bibitem[{{Reames}(1999)}]{reames1999}
{Reames}, D.~V. 1999, \ssr, 90, 413

\bibitem[{{Reames} {et~al.}(1988){Reames}, {Dennis}, {Stone}, \&
  {Lin}}]{reames1988b}
{Reames}, D.~V., {Dennis}, B.~R., {Stone}, R.~G., \& {Lin}, R.~P. 1988, \apj,
  327, 998

\bibitem[{{Reames} \& {Stone}(1986)}]{reames1986}
{Reames}, D.~V. \& {Stone}, R.~G. 1986, \apj, 308, 902

\bibitem[{{Reames} {et~al.}(1985){Reames}, {von Rosenvinge}, \&
  {Lin}}]{Reames1985}
{Reames}, D.~V., {von Rosenvinge}, T.~T., \& {Lin}, R.~P. 1985, \apj, 292, 716

\bibitem[{{Schrijver} \& {De Rosa}(2003)}]{schrijver2003}
{Schrijver}, C.~J. \& {De Rosa}, M.~L. 2003, \solphys, 212, 165

\bibitem[{{Temerin} \& {Roth}(1992)}]{temerin1992}
{Temerin}, M. \& {Roth}, I. 1992, \apjl, 391, L105

\bibitem[{{Wang} {et~al.}(2006){Wang}, {Pick}, \& {Mason}}]{wang2006ApJ639}
{Wang}, Y.-M., {Pick}, M., \& {Mason}, G.~M. 2006, \apj, 639, 495

\bibitem[{{Yokoyama} \& {Shibata}(1996)}]{Yokoyama1996}
{Yokoyama}, T. \& {Shibata}, K. 1996, \pasj, 48, 353

\end{thebibliography}
\end{document}